\documentclass[aps,twocolumn,prl]{revtex4}
\input{epsf}
\newcommand{\be}{\begin{eqnarray}}
\newcommand{\ee}{\end{eqnarray}}

\begin{document}

\title{The patch like model of galaxies formation: the virial
  paradox and core--cusp and missing satellite problems}

\author{M. Demia\'nski$^{1,2}$}
\address{$1. $Institute of Theoretical Physics, University
of Warsaw,\\ 02-093 Warsaw, Poland\\
$2. $Department of Astronomy, Williams College,
Williamstown, MA 01267, USA\\}
\author{A. Doroshkevich$^{3,4}$}
\address{$3. $Astro Space Center of Lebedev Physical
           Institute of  Russian Academy of Sciences,
                        117997 Moscow,  Russia\\
        $4. $National Research Center Kurchatov Institute,
           123182, Moscow, Russia\\}
\date{\today}

\begin{abstract}

The patch like model of the hierarchical galaxy formation in
the $\Lambda$CDM cosmological model with small damping scale
is considered. In this model galaxies and clusters of galaxies
are identified with rare high density peaks, what suppresses
the action of random factors in the vicinity of peaks and
makes the process of halos formation more rapid and regular.
High concentration of irregular subhalos surrounding the
central peaks and their subsequent merging just after formation
allows to consider this medium as a mixture of collisionless
dispersed dark matter (DM) particles and collisional subhalos.
Merging of these subhalos with the central dominating halo is
accompanied by tidal destruction
of the central cusp, what progressively shallows the density
profile and promotes formation of super massive central black
holes. The simulations \cite{ishi14, angulo17,delos19} provide
some quantitative characteristics of these processes.

In the framework of this model we can reproduce the observed
correlation of mass and density of virialized galaxies and
clusters of galaxies known as the {\it virial paradox}
\cite{qsr20,ddlp20}. These correlations are closely linked
with the composition of DM and the shape of the power spectrum
of density perturbations what allows to restrict these
important properties with
already available observations. In particular, these
correlations put constraints on the HDM
and WDM models and allow to test models of cosmological
inflation. We confirm that the missing satellite problem
is directly linked with the {\it virial paradox} and
reheating of the Universe which increases temperature
and entropy of the baryons, prevents formation of first stars
and divides halos into two populations: the first one includes
galaxies formed before reheating which are mainly concentrated
in the vicinity of the massive ones while the second population
-- numerous dark halos formed after reheating --- accumulates
majority of DM but does not contain stars. Their spatial
distribution is more homogeneous.
\end{abstract}
%%\begin{keywords}
%%cosmology: DM dominated observed and simulated galaxies
%%and clusters of galaxies -- properties and evolution --
%%comparison with PS model
%%\end{keywords}
\maketitle

\section{Introduction}

During the last decade much progress has been achieved in
observations of the cosmic microwave radiation (WMAP and
Planck missions, \cite{koma,ade16}, see also \cite{appl20})
and in simulations of the Large Scale Structure of the Universe,
DM halos, galaxies and clusters of galaxies. Many recent
reviews \cite{mcqu,bull,zasov,naab,Tuml,wechs,salu19,bose19,
sawala,simon19,mart20} present various aspects of these
processes.

Now the main attention is concentrated on evolution of the
baryonic matter and formation of observed luminous galaxies
\cite{white78,white91,Guo,qu,bosh18,salu19,salcido,bahe}.
However a few fundamental problems of galaxy formation
remain unsolved and are now actively discussed. First of
all these are the core--cusp and missing satellites problems.
These problems arise when  one  compares  the
observed galaxies and the present day simulated DM halos.
In the last few years unexpected discovery of the ultra
diffuse galaxies \cite{vandok18}, galaxies with a deficit
of DM component \cite{oman16,guo19} and a progress in
understanding of the Ly-$\alpha$ forest \cite{bolt17,
danforth16,dd19,qsr20} complicates the problem of DM
-- baryons interconnection.

Present day high resolution simulations provide several
representative samples of DM halos ranging from dwarf
galaxies and up to rich clusters of galaxies. In majority of
simulations \cite{ishi14,angulo17,delos19,NFW95,nature20}
the DM density profile near the center of halos can be
characterised as cusp--like and is often approximated
by a simple power law:
\be
\rho(r)\propto \rho_0/r^\alpha,\quad  1.5\geq
\alpha\sim 1\,.
\label{cusp}
\ee
This profile reproduces reasonably well the observed
one in clusters of galaxies. However in less massive
galaxies the density profile is more shallow
\cite{deb,pont,Fern} and  is usually described by
more complex expressions  \cite{nature20}.

This is the core-cusp problem. After 20 years of study
discrepancies between models of CDM universe and
observations of low mass galaxies are still pronounced.

Now many models attempt to explain this problem. It has
been  suggested that the core-cusp conflict could be
resolved by fluctuations of the inner gravitational
potential. The most popular explanation of these
fluctuations relies on sudden removal of gas from centers
of cuspy DM halos caused by energy injected by supernovae
\cite{NFW-96,klyp09}. Progressive disruption of the DM cusp
owing to its tidal interaction with the gaseous clouds,
stars and protostars is discussed in \cite{elzant16,de16,
de17,del}. In \cite{Freun19} erosion of the cusp  is related
to accretion of suitable spherical DM shells.

Of course, these factors lead to some cusp erosion, but
a correct estimate of their efficiency is lacking.
Recent analysis \cite{bose19,benitez19} indicates
insufficient variety of mass profiles to explain the
observed diversity of dwarf galaxy rotation curves. In
turn recent models \cite{Freun19}
require a very special kind of DM accretion and conformity
between masses of the cusp and the shells.

Another possibility is to use a more complex dark matter
model. It remains an open possibility that these tensions
may point to exotic particle physics. Such models -- the
scalar field dark matter, Bose -- Einstein condensate,
or ultralight axion DM \cite{Bernal-18,hui17} are identical
to the CDM model at cosmological scales but differ at
galactic scales. All these problems are also reviewed
in \cite{mart20} with more attention put on exotic
particle physics.

However the simplest and the most promising models of the
cusp disruption were discussed in
\cite{ishi14,angulo17,delos181,delos182,delos19}
where it was shown that the cusp becomes shallower in the
earlier formed low mass DM halos owing to merging of
subhalos. This means that the core--cusp problem is mostly
a result of insufficient resolution of present day
simulations and it disappears in  CDM models with a
sufficiently small dissipative scale.

In the standard  $\Lambda$CDM model
gravitationally bound DM structures build up hierarchically
by a combination of accretion of the diffused surrounding
matter and continuous absorbtion of smaller surrounding halos
\cite{white78,white91,sawala,salcido,qu,mart20,nature20}.
During the period of mildly nonlinear matter evolution
the formation of structure elements is driven by the
random velocity field, and at all redshifts it leads to
significant matter concentration in filaments and sheets
\cite{shand83,shand89,dd11,zeld70,dw04,feld18,shand20,ishi20}.
These elements of the structure represent the intermediate
asymptotic of the matter condensation and are observed as
the Large and Super Large Scale Structure \cite{dw04}.
Later on some fraction of remaining loosely distributed matter is accumulated into
compact halos. %%This step of evolution
Formation of DM halos is well described by the
Press--Schechter model (PS) \cite{sheth02,press74,bond91,
klypin11}

In this paper we reconsider the process of DM halos
formation in the framework of the $\Lambda$CDM model with
the power spectrum of density perturbations with a small
damping scale %%with $L_D\sim 7pc$, $M_D\sim 10^{-6}M_\odot$
\cite{Bert}. In this model galaxies are
associated with the very rare random highest peaks of
density perturbations surrounded by many smaller peaks
in the immediate vicinity of the central one. These
special features lead to a very rapid regular growth of
mass of the main halo and progressive tidal disruption of
both the absorbed subhalos and the central cusp.

This is the patch like process of formation of  massive DM
halos when at high redshifts the active creation of new
halos is concentrated only in the vicinity of the central
peak. Later on at $z\leq 10$ many DM halos are formed in
all of space. This model preserves the main features of
the usually discussed models of galaxy formation and the
large scale matter distribution but the internal structure
of early formed halos is more shallow.

Owing to limited resolution, the early period of halos
formation  at $z\geq 10$ is poorly reproduced by
present day simulations \cite{ishi14,angulo17,delos19,nature20}
and for qualitative
estimates we have to relay on the Press--Schechter approach
\cite{press74,bond91,klypin11,sheth02,delos19}. Such
analysis emphasises the important impact of the shape of
the power spectrum and rapid formation of halos at
high redshifts $z\geq 10$. It also illustrates a very
important role of merging of earlier formed subhalos. Traces
of these processes are seen in some present day simulations
and are discussed in \cite{naab,sawala,drako19}.

As was shown in \cite{ddlp20} for objects with virial masses
$10^6M_\odot\leq M_{vir}\leq 10^{14}M_\odot$ the virial density
is a regular function of the mass. For galaxies this
correlation is traced up to redshift $z\sim 4$ \cite{qsr20},
but it fades for both the observed and simulated low mass
DM halos \cite{qsr20,ddlp20}. This property of DM halos
-- the virial paradox -- is also reproduced by the
considered model and allows to restrict the shape of the
small scale power spectrum. It can be usefull for discussion
of the cosmological inflation and puts restrictions on the
WDM models with light DM particles.

Thus considered here patch like model provides satisfactory
description of the observed Universe and attenuates
differences between properties of the observed and simulated
matter distributions.  It leads to a more shallow internal
structure of halos, introduces differences between galaxies
and later formed dark halos and thereby explains both the
virial paradox and the missing satellite problem.
Simple limited versions of this model have been simulated
in \cite{diemand,ander11,ander13,ishi14,angulo17,delos181,
delos182,delos19}. Further progress can be achieved with
special more refined and representative simulations
\cite{ishi20,nature20}.

This paper is organized as follows: the basic properties
of the PS and Zel'dovich approaches are discussed in Sec.
2\,\&\,3, some aspects of the process of halos formation
in the patch like model are discussed in Sec. 4. Conclusions
can be found in Sec. 5. Statistical characteristics of the
Zel'dovich approach are presented in the Appendix.

\section{2. Basic  parameters of the $\Lambda$CDM model}

The standard $\Lambda$CDM model assumes the isotropic matter
expansion with the Hubble constant $H_0$, adiabatic density
perturbations with the Harrison -- Zel'dovich primordial
power spectrum $P(k)$, the dimensionless densities of dark
energy, $\Omega_\Lambda$, dark matter, $\Omega_{DM}$ and
baryonic matter, $\Omega_b$. The density of
nonrelativistic matter, DM, and baryons together, is
determined as $\Omega_m=\Omega_{DM}+\Omega_b$. Observations
of Planck \cite{ade16} allowed to measure these
parameters with high precision
\[
H^2=H_0^2[(1+z)^3\Omega_m+\Omega_\Lambda],\quad
H_0\simeq 67.8km/s/Mpc\,,
\]
\be
\Omega_\Lambda\simeq 0.72,\quad \Omega_{DM}\simeq 0.24,\quad
\Omega_b\simeq 0.04,\quad \Omega_m=0.28\,.
\label{cmbdd}
\ee
Here $z$ denotes the redshift and  the density of nonrelativistic
matter is
\be
\langle\rho_m\rangle =33(1+z)^3\Theta_mM_\odot/kpc^3,
\quad \Theta_m=\Omega_m/0.28\,.
\label{basic}
\ee
For this model the growth of perturbations in the linear theory
can be approximately described as
\be
D(z\geq 1)\approx \frac{1.3}{1+z}\,.\quad
\label{D1}
\ee
This simple fit \cite{dd04} is reasonably accurate for the more
interesting case $z\geq 1$. More refined expression normalized
by the condition $D(0)=1$ can be found in \cite{klypin11}.

\subsection{2.2 Characteristics of the random density and
velocity fields}

In this paper we consider the power spectrum with the
Harrison -- Zel'dovich asymptotic, $P(k)\propto k$,
at $k\rightarrow 0$, and CDM-like transfer function,
$T^2(k)$, introduced in \cite{BBKS}
\be
P(k)=\frac{A^2}{4\pi}l_0^4kT^2(kl_0)D_w(kl_D)\,,
\label{bbks}
\ee
\[
l_0=\frac{Mpc}{\Omega_{m}h^{2}}\simeq\frac{7.14}{\Theta_m}
Mpc,\quad M_0=\frac{4\pi}{3}\langle\rho_m\rangle l_0^3\simeq
\frac{5\cdot 10^{13}M_\odot}{\Theta_m^{2}}\,.
\]
\be
T(x)=ln(1+2.34x)/2.34x/\epsilon(x)
\label{trans}
\ee
\[
\epsilon^4(x)=1+3.89x+(16.4x)^2+(5.46x)^3+(6.71x)^4
\]
Here $k$ is the comoving wave number and $A$ is the
dimensionless amplitude of perturbations. The damping
function $D_w$ and the damping scale, $l_D$, describe
damping of perturbations owing to the random motions
of DM particles. According to \cite{Bert} (see also
\cite{loeb05}) for DM particles with mass $M_{DM}=100GeV$
the damping mass $M_D$ can be taken as
\[
M_D\simeq\frac{4\pi}{3}\langle\rho_m\rangle l_D^3\sim
10^{-5}M_\odot\,.
\]
The damping functions $D_w$ depends on properties of the
DM particles. Here we assume that the spectrum terminates
at $kl_0\simeq 10^{-4}$, $M_D\simeq 50M_\odot/\Theta_m^2$.

For the power spectrum (\ref{bbks}) the dispersion of
the density perturbations is divergent and it is
measured in units of $\sigma_8$ which is the relative
density perturbation, $\delta\rho/\langle\rho\rangle$,
in a sphere of radius $R_8=8h^{-1}Mpc=1.6l_0$,
\be
\sigma_8^2=\int_0^\infty d^3kP(k)W^2(R_8k)\approx
\frac{A^2}{236}=0.64\,,
\label{sig8}
\ee
\[
A\approx 12,\quad W(x)=3(\sin x-x\cos x)/x^3\,.
\]
Here, $W(Rk)$ is the Fourier transform of the real space
top-hat filter corresponding to a sphere of radius $R$ and
mass $\mu=M/M_0$.

In this model the amplitudes of random velocity, $\sigma_u$,
and random displacement, $\sigma_s$, are
\[
\sigma_s^2=\int_0^\infty d^3kP(k)/k^2\approx (1.8l_0)^2,
\]
\be
\sigma_s\simeq 13Mpc,\quad\sigma_u=H_0\sigma_s\simeq
900km/s\,.
\label{sig-s}
\ee

\section{3. The Press--Schechter and Zel'dovich models of
  structure formation}
\label{sec2}

The most popular description of evolution of
perturbations is the linear theory, discussed in many
publications. Unfortunately nonlinear studies of matter condensation
can be described analytically only for special cases. In
spite of the limited applicability of these models they
allow to describe and illustrate the action of some factors
that are important for the structure formation and evolution.

\subsection{3.1 The extended Press -- Schechter model}

Evolution of spherical compact high density objects -- DM
halos, galaxies and clusters of galaxies -- can be
approximated by the  Press--Schechter (PS) model
\cite{press74,bond91,sheth02}.  This model  considers the
successive spherical halo formation around random density
peaks. It assumes the Gaussian distribution function for
the masses accumulated in a spherical volume of radius
$R$ with dispersion
\be
\sigma_m^2(\mu)=\int_0^\infty d^3kP(k)W^2(kR),\quad
\mu=\frac{4\pi\langle\rho_m\rangle R^3}{3M_0}\,.
\label{ps}
\ee
Formation of halos is determined by the condition
\be
D(z)A_{rnd}\sigma_m(M)=1.686,\,\,\, 1+z\simeq
0.77 A_{rnd}\sigma_m(\mu)\,.
\label{zps}
\ee
Here the function $D(z)$ is given by (\ref{D1}) and
$A_{rnd}\geq 0$ characterizes the random height of separate
peaks. Its distribution is described by the Gaussian function
\[
 dW=\sqrt{2/\pi}\exp(-A_{rnd}^2/2)dA_{rnd}\,.
\]

For the spectrum (\ref{bbks}) with the transfer function
from (\ref{trans}) and $M_D\leq M_\odot$ the important
function $\sigma_m(\mu)$ can be fitted by the expression
\be
\sigma_m(\mu)\simeq \frac{3\mu^{-0.06}}{1+1.82\mu^{0.24}}\,.
\label{sigma_m}
\ee

The redshift evolution of the fraction of compressed
matter, $f_m(z,M)$, is given by \cite{sheth02,klypin11}
\be
\frac{df_m(z,M)}{dM}=0.37\frac{dy}{dM}\exp(-y^2)[1+
0.81/y^{0.6}]\,,
\label{fmass}
\ee
\[
f_m(z,M_{min})=0.18\Gamma(0.5,y_{mn})+0.144\Gamma(0.2,y_{mn})\,,
\]
\[
y(z,M)\simeq 1./D(z)/\sigma_m(\mu),\quad y_{mn}=y(z,M_{min})\,.
\]
Here $M_{min}\leq M\leq \infty$, $\Gamma(\beta,x)$ is the
incomplete gamma function. For
$y_{mn}\simeq 0.25(1+z)\mu_{min}^{0.06}\ll 1$, we get
\be
f_m\simeq 1-0.27y_{mn}-0.65y_{mn}^{0.4}+..
\label{fit_m}
\ee
For small $M_{min}$ this fit correctly describes the function
$f_m$ for $1+z\leq 10$.
For the mean mass of a halo we have
\be
\langle M(z,M_{min})\rangle=\frac{1}{f_m(z,M_{min})}
\int_{M_{min}}^\infty Mdf_m\,.
\label{meanmass}
\ee

\subsection{3.2 The Zel'dovich model}

The first analytic theory of structure formation
was provided by the Zel'dovich approximation
\cite{zeld70,shand83,shand89}. This theory correctly
describes  the early anisotropic stage of matter
condensation and formation of elements of the Large
Scale Structure (LSS) of the Universe -- network of
filaments and walls--Superclusters (Zel'dovich pancakes)
\cite{shand83,shand89,dw04,dd04,dd11,feld18,shand20}.
At all redshifts these elements are formed in the
course of mildly nonlinear self similar process of
matter condensation.

In the Zel'dovich approximation the Eulerian, $r_i$, and
the Lagrangian, $q_i$,  coordinates of particles
(fluid elements) and their velocities are related by
\be
r_i =(1+z)^{-1}[q_i - D(z)S_i({\bf q})]\,,
\label{eq1}
\ee
\[
v_i =dr_i/dt=H(z)(1+z)^{-1}[q_i -D(z)\beta(z)S_i({\bf q})]\,,
\]
\[
\beta(z)=1-d\,lnD(z)/d~ln(1+z)\,.
\]
Here the Lagrangian coordinates of a particle, $q_i$,
are its unperturbed coordinates in the real space,
$r_i(z=0) =q_i$, $v_i$ is its velocity,
and the random vector $S_i({\bf q})$ characterizes the
displacement of a particle from its unperturbed position. The
function $D(z)$ is given by (\ref{D1}).

The statistical aspects of this theory had
been discussed in \cite{d70,shand83,dd99,dd04,dw04,dd11}
and are briefly presented in the Appendix.
The random displacement
of a particle $S_i({\bf q})$ is described by the Gaussian
distribution function with the correlation functions
\be
\Psi_{ij}(q)=\frac{\langle S_i({\bf q_1})S_j({\bf q_2})
  \rangle}{\sigma_s^2}=\int_0^\infty dk\frac{k_ik_j}{k^2}
\frac{P(k)W(kq)}{\sigma_s^2}\,,
\label{sij}
\ee
\[
\Psi_{ij}(q)=\frac{1}{3}\delta_{ij}G_1(q)+
\frac{q_iq_j}{3}G_2(q),\quad
\Psi_{ij}(0)=\frac{1}{3}\delta_{ij}\,,
\]
\[
G_1(q)=\frac{4\pi}{\sigma_s^2}\int_0^\infty dk W(qk)P(k),
\quad G_2(q)=\frac{dG_1(q)}{qdq}\,.
\]
Here $q_i=({\bf q_1}-{\bf q_2})_i$\,,
$q=|{\bf q_1}-{\bf q_2}|$, $W(x)$ is the filter
function (\ref{sig8}), $\sigma_s$ is the dispersion of
displacement (\ref{sig-s}). For the power spectrum
(\ref{bbks}) these functions are fitted by
\[
g_{12}=\frac{G_1(q)-1}{q^2G_2(q)}\simeq 0.5(1+0.43\mu^{1/4}+
0.46\mu^{0.04})\,,
\]
\be
G_2(q)\simeq -\frac{0.546\mu^{-0.127}}{1+
  3.86\mu^{0.34}}=-\frac{\sigma_m^2}{5}G_{sm}^2\,,
\label{G2}
\ee
\[
G_{sm}\simeq 1+0.16\mu^{0.17}+0.16\mu^{0.03}\,,
\]
where $\sigma_m$ and the dimensionless mass $\mu$ are given
by (\ref{ps}). These relations emphasize the close link
between the evolutionary rate in the Zel'dovich theory
and the PS model.

The Zel'dovich approach emphasizes the strong impact of
the anisotropic compression and allows to characterize
evolution of the Large Scale Structure rather then formation
of distinct DM halos.

\begin{figure}
\centering
\epsfxsize=8.cm
\epsfbox{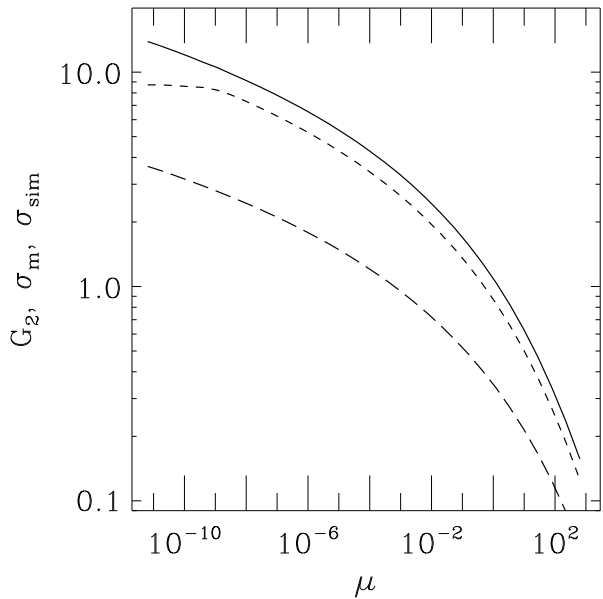}
\vspace{-0.4cm}
\caption{For spectrum (\ref{bbks}) the functions $\sigma_m$,
with $\mu\geq 10^{-11}$, and $\mu\geq 10^{-9}$, and function
$\sqrt{-G_2}$ are plotted vs. dimensionless mass $\mu$
by solid, dashed and long dashed lines.
}
\label{fmz}
\end{figure}

\subsection{3.3 Angular momentum of DM halos}

The Zel'dovich theory predicts an asymmetrical collapse
what decreases the matter compression and generates the
angular momentum for both the separate particles and for
halos as a whole. This problem was discussed in
\cite{ttt,d70,wtt,dd04,dd11,dev19,weyg18,weyg19,weyg20}.

The corresponding components of the velocity are
characterized by the functions
\be
u_i({\bf q})=H(z)e_{ijk}q_jS_k({\bf q})\,,
\label{mom1}
\ee
\[
H^{-2}\langle u_1({\bf q})u_1({\bf p})\rangle=
(q_2p_2+q_3p_3)G_1+(q_2p_3-q_3p_2)^2G_2\,,
\]
\[
H^{-2}\langle u_1({\bf q})u_2({\bf p})\rangle
=-q_2p_1G_1+(q_1p_3-q_3p_1)(q_3p_2-q_2p_3)G_2\,,
\]
\[
G_1=G_1(|{\bf p-q}|),\quad G_2=G_2(|{\bf p-q}|)\,,
\]
and similar relations for other indexes. The angular
momentum of one particle is described as
\be
{\bf p=q},\quad p\ll l_0,\quad u^2(q)\propto 2q^2\,.
\label{mom3}
\ee
The angular momentum of a halo is mainly determined
by an even part of the functions (\ref{mom1}) and we get
\be
J_1^2\simeq H^2\int d^3qd^3p(q_2p_3-q_3p_2)^2G_2\,.
\label{J123}
\ee

\section{4. Patch like galaxy formation in the
  $\Lambda$CDM model}

In the standard  $\Lambda$CDM model a gravitationally bound
DM structure is built up hierarchically by a combination of
accretion of the diffuse surrounding matter and sequential
merging of subhalos \cite{white78,white91,sawala,salcido,
nature20,qu,drako19}. In the $\Lambda$CDM model with a small
damping scale galaxies can be identified with the very rare
high density peaks while between them the low amplitude
perturbations evolve in the linear regime. This difference
results in a patch--like galaxy formation which is operating
up to small redshifts. It is observed as concentration of
denser low mass satellites around massive galaxies and
domination of less dense and massive dark DM halos in
voids.

The massive halos accumulate many less massive subhalos,
which could contain in turn smaller subhalos. High
concentration of subhalos near the highest peaks accelerates
their merging as compared with accretion of the dispersed
DM particles. In turn tidal interactions of the merged loose
subhalos with the central cusp of the major halos leads
to destruction of subhalos, flattening of the cusp
\cite{ishi14,angulo17,delos182,delos19,drako19} and facilitates
formation of super massive black holes. Efficiency of
these random processes depends upon the peak amplitude: it
is high at high redshifts and decreases with time. On the
basis of the present-day simulations the process of halos
formation, the important role of subhalos, their tidal
disruption and heating etc. are discussed in
\cite{naab,sawala}.

Unfortunately, only very complex modern  simulations
\cite{nature20} can describe evolution of a box $L\geq 10Mpc$
with mass resolution $M_{min}\leq M_\odot$ but with only moderate
number of halos formed at redshifts $z\geq 10$. Thus to
study evolution of DM halos we have to rely on the
Press--Schechter formalism and compare its predictions with
numerical simulations \cite{ishi14,angulo17,delos19}.

\subsection{4.1 Characteristics of relaxed halos}

In spite of active discussions the adequate description of
the violent relaxation of DM halos is not yet available. Detailed
dynamical analysis of this process was performed in
\cite{fillmore84,gurev95} for slightly perturbed spherical
clouds. The density profile (\ref{cusp}) with
\[
\alpha\simeq 1.8 - 2\,,
\]
was found in both publications. However this result has
a very limited applicability as the
spherical collapse is very rare \cite{dd11}. At high
redshifts the simulations \cite{ishi14,angulo17,
delos182,delos19} prefer the density profile with
$\alpha\simeq 1.5$.

Efficient method of identification of  the distinct
virialized elliptical halos has not been proposed yet. The popular phenomenological
description of the virial density
\be
\rho_{vir}\simeq 18\pi^2\langle\rho(z)\rangle\simeq
6.6\cdot 10^3(1+z)^3M_\odot/kpc^3\,,
\label{relax}
\ee
provides rough estimate for spherical systems, but
it overestimates $\rho_{vir}$ for the most abundant
elliptical systems.

The estimate (\ref{relax}) is based on two assumptions.
Firstly, according to \cite{pbs67,zk70}, collapse of
a spherical dust cloud at rest of radius $R_0$ and density
$\rho_0$ with conservation of  mass,
$M$, and energy, $E$, results in a virialized state
\[
E=-\frac{3}{5}G\frac{M^2}{R_0} =\frac{1}{2}U=
-\frac{3}{5}G\frac{M^2_{vir}}{2R_{vir}}\,.
\]
Here $U$ is the potential energy of the halo and, therefore,
\be
R_{vir}=R_0/2,\quad \rho_{vir}=8\rho_0\,.
\label{vir}
\ee
Secondly, at the moment of collapse of the homogeneous
spherical halo its average density exceeds the mean
cosmological density by a factor
\[
\rho_0/\langle\rho(z)\rangle=4.5\pi^2.
\]
\cite{lacey93} what in combination with (\ref{vir})
results in expression (\ref{relax}).

This estimate remains correct for the Tolman model of
evolution of a spherical dust cloud \cite{ll75}, but
for deviations from spherical symmetry, such as
ellipsoidal deformations \cite{bona15,des17,balest16,bona18},
the energy $E$ decreases with compression of DM.
The kinetic energy of rotation and turbulent motions also
decrease the expected density (\ref{relax}). Thus the
expression (\ref{relax}) can be considered only as {\it an
approximate phenomenological estimate of the complex process
of relaxation of collissionless DM}.  Owing to its
approximate character the coefficient $500$ is often used
in expression (\ref{relax}) instead of the model
coefficient $18\pi^2$.

Moreover the expression (\ref{relax}) leads to an unexpected
inference that the virial density depends only upon redshift
and, therefore, it is the same for all halos at a
given redshift. The wide variety of observed and simulated
relaxed halos at $z\leq 1$ indicates that the actual
situation is more complex, and it is necessary to restrict
such universality. For this purpose it is convenient to
introduce the redshift of halo formation, $z_{cr}$
\cite{sheth04} and to use (\ref{relax}) with $z_{cr}$. Such
modified version of (\ref{relax}) agrees with the model of
Lacey and Cole \cite{lacey93}. Thus the noted above
replacement of factors $18\pi^2$ by $500$ implies introduction
of redshift of halo formation as
\[
(1+z_{cr})^3=2.5(1+z)^3\,.
\]

Next problem is the correct determination of the shape and
boundary of galaxies and clusters \cite{dd11,bona15,des17,
balest16,bona18,more16,libes18,sheth04} and determination
of halos density in observations and simulations. Thus, very
detailed analysis of halos evolution \cite{klypin11} is
performed without consideration of possible anisotropy of
matter distribution. This factor is specially important for
earlier halos, which often resemble  to flattened
ellipsoids (Zel'dovich pancakes) \cite{dd11}. Dependence of
the halo parameters upon its internal structure \cite{despali17,
connor18,bonamig18} and/or complex environment \cite{umehata19}
should be discussed separately (see, e.g. \cite{salu19}).

\subsection{4.2 DM halos as counterparts of galaxies}
\label{counter}

Some information about the process of galaxy
formation can be obtained using the standard technique
developed for the description of evolution of random density
and velocity fields \cite{shand83,BBKS,shand89,dd99,dd04,
dw04,dd11,feld18,delos19}. In order to identify density peaks with
galaxies and clusters of galaxies we can compare their
mean number densities $\langle n_{cls}\rangle$ and
$\langle n_{pk}\rangle$.

In the SDSS for the observed clusters of galaxies with
$M_{cls}\geq 10^{13}M_\odot$ the mean number density
$\langle n_{cls}\rangle$ and the mean cluster separation,
$\langle d_{cls}\rangle$ are estimated as \cite{vik09}
\be
\langle n_{cls}\rangle\sim 10^{-5}Mpc^{-3},\quad
\langle d_{cls}\rangle\sim 45Mpc\sim 6l_0\,.
\label{ngal}
\ee
In turn for the power spectrum (\ref{bbks}) the mean
number density of high peaks is determined by the scale $l_0$.
This means that clusters are associated with only a small
fraction of the high density peaks.

As was shown in \cite{BBKS} the cumulative number
density of high peaks of a random scalar field can be
roughly estimated as
\be
\langle n_{pk}\rangle\simeq \frac{10^{-2}}{l_0^3}
\exp\left(-\frac{A_{rnd}^2}{2}\right)\simeq n_{cls}\,,
\label{ndens}
\ee
where $A_{rnd}$ is the peak amplitude.
This means that clusters with parameters (\ref{ngal}) are
identified with the peaks of amplitude
\be
A_{rnd}\sim \sqrt{2~ln(\langle n_{pk}\rangle/\langle
n_{cls}\rangle})\simeq 1.4\,.
\label{nu}
\ee
For comparison, assuming cosmological origin of Super
Massive Black Holes, we can estimate the corresponding
peak amplitude for $M_{BH}\sim 10^{10}M_\odot$ as
 \cite{kelly10,decarli20,onken20}
\be
 n_{BH}\sim 3\cdot 10^{-9}Mpc^{-3},\quad A_{rnd}\simeq 4.2\,.
\label{smbh}
\ee

For galaxies the random amplitude $A_{rnd}$ can be
estimated from (\ref{zps}) and the assumption that the
reionization of the Universe at redshifts $1+z\sim 10$
is caused by halos with mass
$M\sim 10^6M_\odot, ~~\mu\sim 2\cdot 10^{-8}$.
These parameters correspond to $\sigma_m(\mu)\sim 0.7 - 0.8$
and thus $A_{rnd}\sim 1 - 1.2$. More accurate estimate of the
peak's amplitude associated with galaxies --
$A_{rnd}\simeq 1 - 2$ will be given in the next subsection.

Of course, this picture predicts formation of many low
mass halos in the vicinity of the main halo in a
wide range of scales. This successive formation of many
low mass DM subhalos continues up to small redshifts. In
turn, large scale perturbations regulate the spatial
distribution of low mass halos, provide their higher
concentration in the immediate vicinity of the central peak
and regulate the further transformation of the system of
subhalos into the distinct massive objects -- galaxies,
filaments or sheet--like superclusters.

\subsection{4.3 The virial density of galaxies and clusters
  of galaxies: the virial paradox}

As was discussed in \cite{ddlp20} for galaxies and clusters
of galaxies %%with masses $10^6\leq M/M_\odot\leq 10^{14}$
the virial density, $\rho_{vir}$, is a regular function of
their mass. Thus for the sample of 194 observed galaxies and
447 clusters of galaxies with
$10^6\leq M_{vir}/M_\odot\leq 10^{14}$ the reduced virial density
$G_\rho$ is fitted as follows:
\be
G_\rho^{obs}(\mu)=\frac{\rho_{vir}\sqrt{\mu}}{\langle G_\rho\rangle}
\simeq\frac{40\mu^{0.25}}{(1+2\mu^{0.25})^3}\,,
\label{obs}
\ee
\[
\langle G_\rho\rangle=6\cdot 10^5M_\odot/kpc^3
\simeq 2\cdot 10^4\langle\rho_m(0)\rangle\,.
\]
The maximal value $G_\rho^{obs}\sim 3$ is achieved for
$\mu\sim 4\cdot 10^{-3}$, $M\sim 2\cdot 10^{11}M_\odot$.
In Fig. \ref{fmz} the function $G_\rho^{obs}$ is plotted vs.
virial mass of halos $M_{vir}/M_\odot$. More detailed analysis
of several clusters \cite{bona15,des17,balest16,bona18}
confirms this estimate \cite{ddlp20}.

\begin{figure}
\centering
\epsfxsize=8.cm
%%\epsfbox{epsfiles/cls_gal_th.eps}
\epsfbox{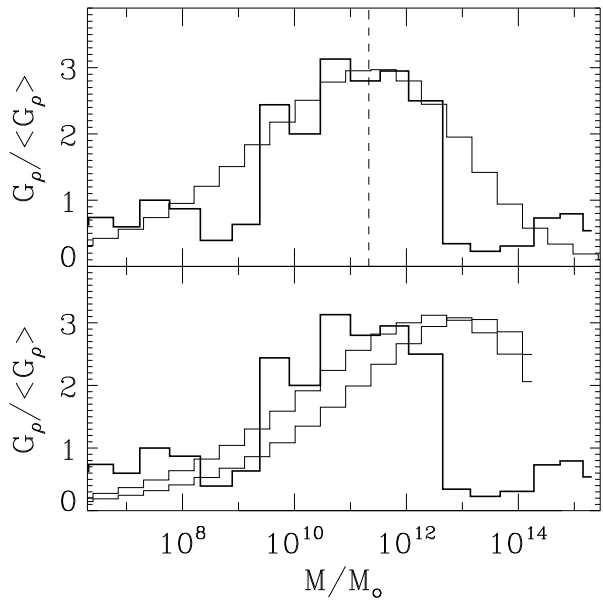}
\vspace{-0.4cm}
\caption{Top panel: the observed PDF of reduced virial
density $G_\rho^{obs}(\mu)$ (\ref{obs}) is plotted by solid
line vs. virial mass of halo $M/M_\odot$. Theoretical function
$G^{th}_\rho(\mu)$ (\ref{gr507}) is plotted by thin line, dashed
line marks the position of maximum. Bottom panel: the same
observed PDF $G_\rho^{obs}$ is compared with the PDFs
(\ref{r402}) and PDF, for function $\sigma_m(M)$ presented in
\cite{klypin11}).
}
\label{fmz}
\end{figure}

The corresponding theoretical function can be found with
Eqs. (\ref{zps}), (\ref{sigma_m}) and (\ref{relax})
\be
G_\rho^{th}(\mu)=2.3\cdot 10^{-2}(A_{rnd}\sigma_m)^3
\sqrt{\mu}\,,
\label{thr}
\ee
and it is closely linked with the function $\sigma_m(\mu)$
and the power spectrum (\ref{bbks}). For the standard power
spectrum \cite{BBKS}  $\sigma_m(\mu)$ is given by
(\ref{sigma_m}) and we get from (\ref{thr})
\be
G_\rho^{th}(\mu)\simeq \frac{0.6A^3_{rnd}\mu^{0.32}}{(1+
1.8\mu^{0.23})^3}\,.
\label{r402}
\ee

As is shown on Fig. \ref{fmz}, the maximal value of
the function (\ref{r402}) is achieved for $\mu_{max}\sim
0.05 - 0.1$, $M_{max}\sim (1-5)\cdot 10^{12} M_\odot$
and required amplitude is $A_{rnd}\simeq 3.5$. Similar
results are obtained for the function $\sigma_m(\mu)$
presented in \cite{klypin11}. These functions are
plotted in Fig. \ref{fmz} (bottom panel).

Differences between the observed function
$G_\rho^{obs}(\mu)$ and theoretical one $G_\rho^{th}(\mu)$
can be reduced by increasing the
observed virial mass of galaxies (by a factor $\sim 10$).
It also can be reduced by deformation of the power
spectrum $P(k)$ (\ref{bbks}). Deformation of both the initial
power spectrum and the transfer function $T(kl_0)$
(\ref{trans}) also can be considered. The large scale power
spectrum is measured by observations of the relic
microwave radiation \cite{koma,ade16}. So, here we
assume that the power spectrum differs from (\ref{bbks})
by a correction function
\be
\psi_{cor}(q)=1+\frac{q^2}{1+a_cq^2},\quad q=kl_0,
\quad a_c\simeq 0.2\,.
\label{correction}
\ee
This function increases the amplitude of the small
scale power spectrum by the factor of $1/a_c$ and retains
its shape. For so corrected power spectrum we have
\be
\sigma_m(\mu)\simeq \frac{5\mu^{-0.066}}{1+3\mu^{0.31}}
\label{sig507}
\ee
and for the reduced virial density
\be
G_\rho^{th}(\mu)\simeq\frac{5A^3_{rnd}\mu^{0.3}}{(1+3\mu^{0.31})^3}\,.
\label{gr507}
\ee
As is seen from Fig. \ref{fmz}  observed (\ref{obs}) and
corrected theoretical function (\ref{gr507}) are similar
to each other for the amplitude $A_{rnd}\sim 2$.

This approach is more sensitive for the small scale power
spectrum but its reliability is yet not very high. Indeed, it is
based on a limited statistic, observations are performed with
limited precision, the theoretical base is the phenomenological
PS model. Non the less these results confirm that for galaxies
and clusters of galaxies the correlation of the virial density
and the virial mass can be a natural result for the corrected
power spectrum (\ref{bbks},\ref{correction}). One would expect
that this approach can be used as a test of models of inflation
\cite{infl14,bal20}, and, in particular, it favors more complex
models of inflation. It also allows to restrict parameters
of WDM models and properties of hypothetical exotic DM particles
\cite{Bernal-18,hui17,mart20}.

Similar links between the halos masses and sizes are found
for 160 systems of metal lines observed in absorption spectra
of quasars \cite{qsr20}, for 30 massive clusters in simulations
\cite{armit18} and for three clusters of galaxies at
$z\simeq 0.4$ \cite{bona18}. However for simulated low
mass halos the reduced virial density is much smaller than
(\ref{obs}). Thus for both halos associated with the $10^3$
Ly-$\alpha$ absorbers \cite{qsr20} and $\sim 10^6$ low
mass simulated halos \cite{ddlp20} we get that
\[
\langle G_\rho^{sim}\rangle\leq 2\cdot 10^{-2}\,.
\]

Simulations demonstrate \cite{ddlp20} that for redshifts
$z\leq 10$ the reduced density $G_\rho$ of low mass halos
is a many--valued function of the mass. This means that either
these halos are not virialized, or the virial density is not
described by the function (\ref{relax}) and, as it was discussed
in Sec. 4.1, we need to use a more complex correction function,
for instance, by inclusion $z_{cr}$ -- the redshift of halos
formation. After such correction the many--valued character
of the simulated virial density gets  reasonable explanation:
halos with the same $\mu$ are formed  with the same
$A_{rnd}D(z_{cr})$ (\ref{zps},\ref{mpeak},\ref{m507})
but different $z_{cr}$,
$\rho_{vir}$ and $G_\rho$. These expectations can be easy
tested in simulations. For observed galaxies and clusters of
galaxies such ambiguity is not so evident owing to the limited
number of observed halos.

This unexpected different properties of observed and
simulated halos can be formulated as the {\it virial paradox}.
It shows that there are at least two populations of DM halos
with different evolution and different properties. One of them
is formed at high redshifts $z\geq z_{thr}\simeq 10$ and is
observed as galaxies, and the other is formed at small
redshifts, $z\leq z_{thr}$ do not contain stars and is
observed, in particular, as Ly-$\alpha$ forest and circum
galactic halos \cite{qsr20}. This effect is closely linked
with the {\it missing satellite} problem.

\subsection{4.4 Evolution of massive halos}
\label{mass-z}

In the PS model formation of halos is determined by the condition
(\ref{zps}) and usually the model is applied to describe
evolution of the mass function (\ref{fmass}) and the mean
mass of halos (\ref{meanmass}).

\begin{figure}
\centering
\epsfxsize=8.cm
\epsfbox{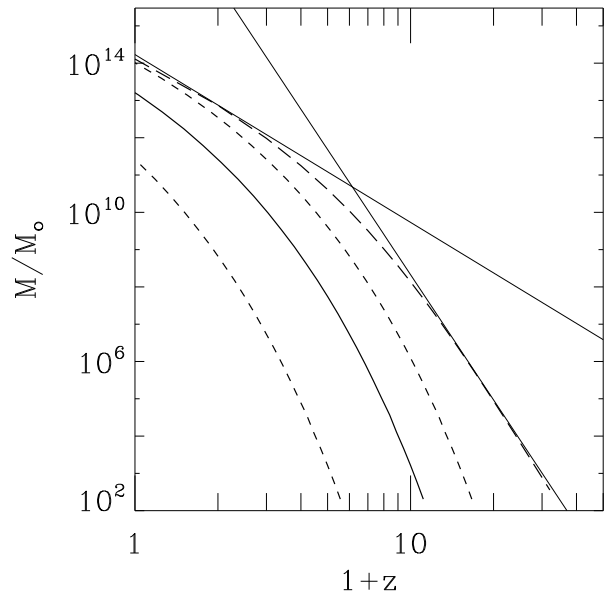}
\vspace{-0.3cm}
\caption{The mass of DM halo (\ref{mpeak}) for amplitudes
$A_{rnd}=0.5, 1, 1.5$ are plotted vs. redshift by dashed
and solid lines. Mean mass of the DM halos
$\langle M(z,M_{min})\rangle$ (\ref{<M>}) is plotted
by long dashed line. Thin straight lines plot the power
fits (\ref{<M>}).
}
\label{<mmm>}
\end{figure}

For the minimal mass $M_{min}=10M_\odot$ we get for the mean mass
\be
\langle M(z)\rangle=\frac{5.5\cdot 10^8M_\odot\,\,
x_{10}^{11.5}}{1+0.5x_{10}^{6.5}+1.8x_{10}^{0.5}},\quad
x_{10}=\frac{10}{1+z}\,.
\label{<M>}
\ee
This function is plotted in Fig. \ref{<mmm>}. At small
redshifts $1\leq z\leq 10$ the mean mass of halos is

\be
\langle M\rangle=1.7\cdot 10^{14}M_\odot
\frac{1\pm 0.2}{(1+z)^{4.5}}\,.
\label{lowz}
\ee

The mass of a separate halo with the random peak amplitude
$A_{rnd}$ can be found with Eqs. (\ref{zps}) and (\ref{sigma_m})
\be
\mu\simeq\frac{1.6\cdot 10^{-4}x^{17.3}}{1+12x^7+x^{11.8}},
\quad x=\frac{7A_{rnd}}{1+z}\,.
\label{mpeak}
\ee
For $z\gg 1$,  $x\ll 1$ this expression is quite similar
to (\ref{<M>}) and for $x\geq 1$ it is similar to
(\ref{lowz}).  As is seen from (\ref{mpeak}) for $x\ll 1$
we have
\be
\mu\simeq \left[\frac{4A_{rnd}}{1+z}\right]^{17}\,.
\label{m507}
\ee

In these expressions the crucial role of the redshift
$z\simeq 10$ is clearly seen since then the
rate of matter concentration in the massive halos strongly changes. In turn,
this change of the evolutionary rate is the direct result
of the complex shape of the power spectrum (\ref{bbks}),
\be
(kl_0)^3P(kl_0)\propto\left\{
\begin{array}{ll}
  ln^2(kl_0)&kl_0\gg 1\,,\cr
  (kl_0)^4&kl_0\ll 1\,.\cr
\end{array}
\right\}
\label{spcc}
\ee
These results are illustrated in Fig \ref{<mmm>}.

The PS model with a small minimal mass shows that already
during the early period of evolution DM halos can accumulate
significant mass fraction and it rapidly increases with
time. Thus for the minimal mass $M_{min}=10^{-2}M_\odot$ at
$z\simeq 50$ the matter fraction $f_m\sim 0.1$ is already
concentrated in halos with $\langle M\rangle\simeq 0.2M_\odot$,
and at $z\simeq 25$ the matter fraction $f_m\sim 0.3$ is
accumulated in halos with $\langle M\rangle\simeq 10^3M_\odot$.
At $z\simeq 10$ the matter fraction $f_m\sim 0.5$ is
concentrated in halos with $\langle M\rangle\sim 10^8M_\odot$,
but majority of DM is concentrated in less massive halos.
At the same time halos with  masses $M\geq 10^6M_\odot$,
$\langle M\rangle\sim 10^9M_\odot$
accumulate only $f_m\sim 0.1$ but in such halos the first
stars are formed, and they are responsible for reionization
and reheating.

There are five natural consequences of the patch like model of
galaxy formation and identification of galaxies with the
highest density peaks:
\begin{enumerate}
\item{} At high redshifts $z\geq 10$ rapid
($M\propto (1+z)^{-17}$ (\ref{m507}) ) regular growth of
the central halo is accompanied by formation of many subhalos
in the immediate vicinity of the central peak and their
rapid merging.
\item{} During this period densities of the central halo
and merged subhalos are similar  and
the shapes of subhalos are far from spherical \cite{dd11}. These
peculiarities allow to consider this medium as a mixture
of collisionless dispersed DM particles and collisional
irregular DM subhalos.
\item{} Rapid infall of surrounding subhalos into the central
  halo and their tidal disruption just after merging
  partly transforms the dissipationless evolution of the
  dispersed DM into the dissipational one of subhalos.
  Impact of the low entropy baryons amplifies this effect.
\item{} Accretion of the diffuse DM particles leads to
  formation of a cusp-like density profile (see, e.g.,
  \cite{ander11,nature20,NFW95}), but successive merging of
  surrounding DM subhalos rapidly increases the  central
  halo and gradually makes its cusp more and more shallow
  \cite{ishi14,angulo17,delos181,delos182,delos19}.
\item{} Accretion of the collisional fraction -- DM
  subhalos -- increases concentration of DM in the central
  core of halos and promotes formation of massive and
  supermassive  central black holes.
\end{enumerate}

These comments clarify qualitative differences in evolution
of DM halos at high and small redshifts. Weak traces of these
processes are seen in the present day simulations and are
discussed in \cite{naab,sawala}. More accurate quantitative
estimates could be obtained from suitable simulations such
as \cite{dd11,delos182,delos19,nature20}.

For the later period ($z\leq 5$) of evolution growth of
halos mass slows
down (\ref{lowz}) and there is a large gap between
formation of massive clusters of galaxies at $z\simeq 1$
and high density galaxies at $z\geq 2$. In contrast with
the high redshift evolution, collisions and striping
of galaxies are rare, and in the observed clusters both
the cusp like density profile and distinct high density
galaxies could survive.

\subsection{4.5 Evolution of the central cusp}

At present the structure and evolution of the core of DM
halos remains unclear as it can not be described
analytically and in simulations it is investigated at high
and low redshift in \cite{ishi14, angulo17,delos19,nature20,
diemer15,diemer19,diemer20}.

At $z\leq 10$ modern high resolution simulations
provide a set of representative DM halos ranging from
dwarf galaxies and up to rich clusters of galaxies. Their
profile is found to be cusp--like and close to the two
parametric Navarro-Frenk-White (NFW) profile
\cite{NFW95}
\be
\rho_{NFW}(r)\simeq \frac{\rho_0}{x^\alpha(1+x)^2},\quad
x=r/r_0,\quad \alpha=1\,.
\label{NFW}
\ee
This profile reproduces reasonably well that observed for
majority of clusters of galaxies. However in less massive
galaxies the observed density profile is more shallow (Table
\ref{tbl-obs}) and can be approximated by power law with
the index $\alpha\leq 1$. Strong influence of random factors
for galaxies manifests itself as larger scatter of the power
index $\alpha$. A review of observations is presented in
\cite{deb}.  More detailed discussion of this problem
can be found in \cite{salu1,salu2,salu3,pao19,salu19}
and references therein.

Now four models of flattening of the central cusp
are discussed:
\begin{enumerate}
\item{} The most popular explanation is the cusp
destruction owing to sudden removal of gas from the center
of a cuspy DM halo caused by energy injection by
explosions of supernovae \cite{NFW-96}.
\item{} In \cite{elzant16,de16,de17,del} the authors consider
progressive destruction of the DM cusp owing to its tidal
interaction with gaseous clouds, first stars and protostars.
\item{} In \cite{Freun19} the cusp erosion is related to
accretion of a suitable DM spherical shell.
\item{}Models with exotic particle physics are presented
in \cite{Bernal-18,hui17,mart20}.
\end{enumerate}

\begin{table}
\caption{Parameters of observed density profile}
\label{tbl-obs}
\centering
\begin{tabular}{rrll}%10c
\hline
$N_{obj}$&$\alpha_{mn}\leq\alpha\leq\alpha_{max}$&
$\langle\alpha\rangle$&reference\cr
\hline
20 clusters&$0.5\leq \alpha\leq 1.5$&$1.02\pm 0.08$&\cite{mantz}\cr
26 galaxies&~~$0\leq \alpha\leq 1.2$&$0.2\pm 0.2$&\cite{debb}\cr
15 galaxies&~~$0\leq \alpha\leq 1.2$&$0.6\pm 0.35$&\cite{swat}\cr
 7 galaxies&~~$0\leq \alpha\leq 1.2$&$0.29\pm 0.07$&\cite{oh11}\cr
26 galaxies&~~$0\leq \alpha\leq 1.2$&$0.32\pm 0.24$&\cite{o15}\cr
 7 galaxies&~~$0.5\leq\alpha\leq 0.73$&$0.67\pm 0.10$&\cite{adam14}\cr
\hline
\end{tabular}
\end{table}

Of course, these factors influence the density profile but
they are of secondary importance with moderate efficiency
which depends upon thermal evolution of baryons, star
formation or high energy physics. Recent discussion of
this problem \cite{bose19,benitez19} confirms the critical
role of the threshold of stars formation and insufficient
variety of mass profiles. Authors of \cite{bose19,benitez19}
believe that their simulations cannot explain the observed
diversity of galactic rotation curves.

However, these inferences relate to simulations with
the cutoff of power spectrum which cannot represent the
earlier stage of halos formation responsible for the
structure of the central region. It can be expected that
matter compression went through a pancake--like
anisotropic stage and the final density profile is formed
by a complex relaxation processes.

In \cite{dd11} these problems have been analyzed in detail
using simulations at redshifts $0\leq z\leq 3$ and the
Minimal Spanning Tree technique. The main results of this
investigation can be formulated as follow:
\begin{enumerate}
\item{}the shape of a halo with mass $M$ is elliptical
with half axes $a_i$ and the velocity dispersion $w_i$
\be
a_1:a_2:a_3\sim (6:2:1)\sqrt{M},\quad
w_i\propto\sqrt{M}\,,
\label{sim11}
\ee
\item{}in accordance with the tidal torque theory
\cite{ttt,d70,wtt} the angular momentum of clouds can be
approximated by $|{\bf J}|\simeq 0.17|w_i|R_{vir}$
with the exponential PDF.
\item{}the turbulent motions can be
approximated by the angular momentum
$|{\bf j}|\simeq 0.8|w_i|R_{vir}$ with the Gaussian PDF.
\end{enumerate}
These results indicate limited applicability of the spherical
approach \cite{fillmore84,gurev95} and a high influence of
velocities for evolution of the central regions of DM halos.
Some applications of the tidal torque theory are discussed
in \cite{weyg18,weyg19,dev19,weyg20}.

High efficiency of tidal interactions for flattening of the
cusp was demonstrated in \cite{elzant16,de16,de17,del} where
the cusp disruption is explained by absorption and tidal
destruction of a suitable set of stars and protostars.
High efficiency of merging of surrounding DM subhalos and
tidal heating of the central cores are confirmed by
simulations \cite{ishi14,angulo17,delos182,delos19,drako19,
penar19}. These papers illustrate the crucial role of initial
stages of halos formation for correct reproduction of
structure of DM halos and, in particular, for successive
transformation of the central cusp into a core. They
demonstrate that
\begin{enumerate}
\item{} For the low mass DM subhalos the central cusp
is steeper than in the NFW model.
\item{} The central cusp becomes shallower owing to
rapid merging processes as the halos mass increases.
\item{} The simulated mass dependence of the power index
can be  roughly fitted as
\be
\alpha\simeq 0.123\log\left(\frac{M}{10^6M_\odot} \right)\,.
\label{ishi}
\ee
\end{enumerate}
This expression \cite{ishi14} describes both the small
$\alpha$ at $M\sim (10^6 - 10^9)M_\odot$ and $\alpha\sim 1$
for clusters with $M\sim 10^{14}M_\odot$.

The new technique \cite{nature20} allows to combine large
simulated box and large redshift interval and it promise a rapid
progress in investigations of the cusp -- core transformations.

\subsection{4.6 The missing satellites problem}

The missing satellites problem is formulated as a strong
discrepancy between the number of observed satellites of the
Milky Way ($\sim 30 - 40$ at distances $D\leq 1Mpc$) and
the number of simulated DM subhalos around massive galaxies
\cite{klypin2015}. It can be reformulated as high
difference between the matter fraction concentrated in
luminous galaxies and in the DM halos ($\sim 70\%$).
Estimates of \cite{shull} show that only moderate fraction
of baryons,
\be
\Omega_{lum}\simeq 0.20(1\pm 0.2)\Omega_b\,,
\label{lum}
\ee
is concentrated in luminous objects (stars, galaxies,
clusters of galaxies).

%This means that t
The main difference between galaxies and
DM halos is the presence or absence of stars, what reduces
the discussion to the problem of formation of stars -- or
even first stars. The virial paradox discussed in Sec.
4.3 shows that here we have to deal with two different
populations of halos and it is closely linked with the shape
of the primordial power spectrum of density perturbations
(\ref{spcc}). Observations of the ultra diffuse galaxies
\cite{vandok18} suggest that there is continuous transition
between  these populations.

The matter fraction (\ref{lum}) related to galaxies
is comparable with that accumulated by high density massive
halos before reheating of the Universe when the temperature
of the low density baryons was rapidly increasing from
$\sim 1 K$ up to $\sim 10^4K$. (Next problem is the
topology of the reionization bubble network \cite{elbers18}).
The high density fraction of  baryons
kept low entropy and there the first stars could have been formed.
This means that we can consider the stars as the trademark of
such halos. The multitude of low mass DM halos formed later
do not contain neither the low entropy baryons nor stars.

Conversion of DM halos into the observed luminous galaxies
is a very complex multistep process
\cite{wechs,GOH2018,somer18,allen18,Behr18} which can be
described only phenomenologically. These complexities prevent
discrimination between simulated galaxies and
invisible DM halos \cite{sff16}.
Besides, %%recent publications
papers \cite{vandok18,oman16,guo19}
show some unexpected features of dwarf galaxies what
emphasizes again a complex character of their evolution.
The PS formalism reproduces the observed estimate (\ref{lum})
and confirms that the missing satellites problem is
deeply linked with the virial paradox and the
identification of galaxies and simulated DM halos.

As is illustrated in \cite{weisz} %Weisz et al. (2014)
metal production in dwarf galaxies is %%strongly
irregular
and randomized. Present day simulations can reproduce these
processes only phenomenologically with %%additional
special assumptions. This means that the missing satellites
problem requires more adequate simulations with restoration
of the patch like formation of galactic counterparts and the
reheating process.

\section{5. Conclusions}
\label{con}

In this paper we consider evolution of the
$\Lambda$CDM cosmological model with a small damping scale.
In this model the patch like character of halos formation
leads to creation of two different populations of objects.
The first population includes high density halos formed
before reheating in immediate vicinity of high density
peaks identified with galaxies. Such halos contain stars
and low entropy baryons and are observed as galaxies. Low
mass halos of the second population are formed after
reheating and they do not contain stars and low entropy
baryons. Some of them are observed as the Ly--$\alpha$
forest \cite{qsr20} and circumgalactic matter
\cite{danforth16}.

Evolution of the second population is investigated in
many simulations. In contrast, evolution of the first
population is presented only in a few simulations
\cite{ishi14,angulo17,delos181,delos182,delos19,nature20}.
For its description we have to use abilities of analytically
extended Press--Schechter and Zel'dovich approaches. Such
models allow to reveal the main specific features of halos
evolution, to clarify differences in properties of these
populations and to explain the virial paradox, the
core -- cusp and the missing satellite problems.
As discussed in Sec. 4, properties of these two families
of halos are determined by the shape of the initial power
spectrum of density perturbations.

Thus the patch like model demonstrates that:
\begin{enumerate}
\item{} The rapid formation of many subhalos at $z\geq 10$
  in the immediate vicinity of the rare high density peaks
  and their rapid merging just after formation leads to a very
  rapid growth of mass of the central halo
  (\ref{<M>}--\ref{m507}) and Fig. \ref{<mmm>}.
\item{} Tidal interaction of irregular merged subhalos with
  the central cusp makes it more and more shallow, what allows
  to explain the core -- cusp problem and accelerates
  formation of the central black hole. These inferences are
  consistent with simulations
  \cite{diemand,ishi14,angulo17,delos181,delos182,delos19}.
\item{} The patch like model demonstrates strong differences
  between characteristics of populations of galaxies created
  before reheating and numerous population of low mass dark
  DM halos created later. The last population does not contain
  low entropy baryons and stars. This division explains
  the missing satellite problem.
\item{} The path like model  explains the {\it virial
  paradox} -- observed correlations of the density and mass
  of first population of halos -- galaxies and clusters of
  galaxies \cite{qsr20,ddlp20} and links it with the
  initial power spectrum of density perturbations.
\item{} This link allows to estimate the shape of the small
  scale power spectrum (Fig. \ref{fmz}) and to place new
  constrains on the parameters of DM particles, WDM models
  and models of cosmological inflation. This approach deserves
  further refined investigations in both observations and
  simulations.
\end{enumerate}

Traces of these processes are revealed in present day
simulations and are discussed in \cite{naab,sawala}. However
these simulations cannot adequately reproduce both the early
and later periods of structure evolution as well as the
reheating of the intergalactic matter. This means that
for our analysis we have to use theoretical models. Results
obtained in this way cannot be considered as an actual proof
of declared inferences but they reveal new important factors,
actions of which were underestimated in previous discussions.
Further progress can be achieved  with new massive observations
of virial parameters of galaxies and with simulations that
use a new approach \cite{nature20}.

\section*{Acknowledgments}
The work is supported by the scientific group
41-2020 of Lebedev Physical Institute.

We wish to thank the anonymous referee for
valuable suggestions and many useful comments.

\section{Appendix: Statistical characteristics of the
  Zel'dovich theory.}.

The Zel'dovich approximation \cite{zeld70,shand83,shand89}
correctly describes the early anisotropic period of matter
condensation and formation of elements of the Large Scale
Structure of the Universe -- network of filaments
and walls--superclusters (Zel'dovich' 'pancakes')
\cite{shand83,dw04,dd04,feid18,shand20}. At all redshifts
these elements are formed in the course of mildly nonlinear
self similar matter compression described by Eq. (\ref{eq1},
\ref{sij}).

The original Zel'dovich model describes the matter
condensation with the deformation tensor. However, for the
CDM models with a small scale cutoff of the power spectrum
this approach has to be reformulated in terms of
displacements of particles what allows to prevent
singularities at $kl_0\gg 1$ and to  concentrate
more attention on observed and simulated scales.
This requires modification of the statistical description
of these processes. These problems have been discussed in
\cite{shand83,dd99,dd04,dw04,dd11} and are shortly
presented here.

As it follows from (\ref{sij}) for the relative of
displacements $\Delta S_i$ we get
\[
\Phi_{ij}({\bf q_1},{\bf q_2})=\langle\Delta S_i({\bf q_1})
\Delta S_j({\bf q_2})\rangle=
2\Psi_{ij}(0)-2\Psi_{ij}(q_{12})\,,
\]
\[
\Delta S_i({\bf q})=S_i({\bf q})-S_i({\bf -q})\,,\quad
q_{12}=|{\bf q_1-q_2}|\,,
\]
\be
G_{11}=\Phi_{11}(q_1,q_1)=
\frac{2}{3}[1-G_1(2q_1)-4q_1^2G_2(2q_1)]\,,
\label{delts}
\ee
\[
G_{12}=\Phi_{12}({\bf q_1},{\bf q_2})
=-\frac{8}{3}q_1^2G_2(\sqrt{2}q_1),
\quad |{\bf q_1}|= |{\bf q_2}|\,.
\]
These relations and eq. (\ref{G2}) allow us to estimate
the coefficient correlations of the orthogonal displacements
for the spectra discussed in Secs. 3\,\&\,4 as
\be
r_{12}=r_{13}=r_{23}=G_{12}(q_1,q_2)/G_{11}(q_1,q_1)\simeq 2/3\,.
\label{r123}
\ee

\subsection{Structure characteristics of uncorrelated
  distribution function of displacements}

According to Eqs. (\ref{eq1},\ref{sij}) the Zel'dovich
approach describes the matter condensation in compact
clouds, filaments and walls. In particular it allows to
estimate the evolution of matter fractions associated with
-- walls $W_w$, filaments $W_f$, clouds $W_{cl}$ and voids
$W_v$. For illustration we can ignore correlations between
orthogonal displacements $\Delta S_i\Delta S_k,\,\,i\neq k$
and to assume that the distribution function for these
displacements is Gaussian
\be
dW=\Phi(\xi_1,\xi_2,\xi_3)d^3\xi=0.75\exp(-Q)d^3\xi\,,
\label{pdf-0}
\ee
\[
Q=(\xi_1^2+\xi_2^2+\xi_3^2)/2,\quad
-\infty\leq \xi_3\leq \xi_2\leq \xi_1\leq\infty\,,
\]
and $\xi_i=D(z)\Delta S_i/q_i$.

According to Eq. (\ref{eq1}) the matter fractions
accumulated by structure elements are determined by a common
threshold $\zeta\geq 0$ and we get for voids,
\[
\zeta\geq\xi_1\geq \xi_2\geq \xi_3\geq -\infty\quad
W_v=0.125(1+e(\zeta))^3\,,
\]
where $e(\zeta)=erf(\zeta)$ and for the distinct clouds, formed directly from a weakly
perturbed matter
\[
\infty\geq\xi_1\geq\xi_2\geq \xi_3\geq\zeta,\quad
W_{cl}=0.125(1-e(\zeta))^3\,.
\]
The matter fractions accumulated by filaments, $W_f$, and
walls, $W_w$, are determined by similar conditions
\[
\infty\geq \xi_1\geq \xi_2\geq \zeta\geq \xi_3\geq -\infty,
\quad W_f=3W_{cl}\frac{1+e(\zeta)}{1-e(\zeta)}
\]
\[
\infty\geq \xi_1\geq\zeta\geq \xi_2\geq \xi_3\geq -\infty,
\quad W_w=3W_v\frac{1-e(\zeta)}{1+e(\zeta)}\,.
\]

Evidently, $W_w+W_f+W_{cl}+W_v=1$. Thus, for $\zeta=0$
we get
\[
W_w=W_f=3/8,\quad W_{cl}=W_v=1/8\,,
\]
but for $\zeta\geq 0$ the symmetry is destroyed
(Table \ref{tbl-frac}) -- walls and voids accumulate
dominant fraction of matter.

\begin{table}
\caption{Matter fraction accumulated by clouds, filaments,
walls and voids for different threshold amplitude $\zeta$
and coefficient correlations of orthogonal  displacements
$\kappa$}.
\label{tbl-frac}
\begin{tabular}{lrr rrr }%8c
$\zeta$&$\kappa$&$W_{cl}$&$W_f$&$W_w$&$W_v$\cr
\hline
     0&   0   &0.125&0.375&0.375&0.125\cr % 1
   0.5&   0   &0.029&0.197&0.443&0.331\cr  % 8
     1&   0   &0.004&0.063&0.337&0.596\cr  %2
     0&   0.66&0.045&0.455&0.455&0.045\cr  %7
   0.5&   0.66&0.004&0.200&0.597&0.200\cr  % 6
     1&   0.66&1.e-4&0.052&0.488&0.459\cr  %3
\hline
\end{tabular}
\end{table}

However as was shown in the first Zel'dovich paper
\cite{zeld70} the size of high density multistream
regions associated with walls exceeds the size determined
by the condition $\zeta=1$ by a factor $\sqrt{3}$.
This means that the condition $\zeta\simeq 0.5$ more
correctly describes the compressed matter fractions.

\subsection{Structure characteristics of the correlated
  distribution functions of displacements}

The impact of correlations of the orthogonal
displacements leads to more cumbersome estimates and can
be suitably analyzed numerically using $10^7$ random
realizations. The correlation coefficient $\kappa$ depends
upon the power spectrum and for the power spectrum
(\ref{bbks}) $\kappa\sim 2/3 $. For three amplitudes
$\zeta$ the matter fractions accumulated by structure
elements are presented in Tabl \ref{tbl-frac}

As is seen from this Table the matter fraction directly
accumulated by clouds, $W_{cl}$ is minimal, but for the
amplitude $\zeta=0.5$ the matter fraction accumulated by
high density LSS elements
, $1-W_v$, increases up to
67\% what is comparable with simulated results.
As was shown in  \cite{sz72} and \cite{d80}  after
formation of high density filaments and walls they are
rapidly disrupted into  distinct relaxed clouds. This
inference is also confirmed by simulations.

In the general case the PDF of the random displacements
is Gaussian with  correlation coefficients $r_{ij}$,
$i,j=1, 2, 3$,
$-\infty\leq \xi_3\leq\xi_2\leq\xi_1\leq\infty$
\be
Q=\frac{\xi_1^2+\xi_2^2+\xi_3^2}{2}+\kappa_{12}\xi_1\xi_2+
\kappa_{13}\xi_1\xi_3+\kappa_{23}\xi_2\xi_3\,,
\label{gauss1}
\ee
\[
\xi_i=\varepsilon_{jk}\Delta S_i,\quad
\kappa_{ij}=\frac{r_{ij}-r_{ik}r_{jk}}{\sqrt{(1-
r^2_{ik})(1-r^2_{jk})}},\quad i\neq j\neq k\,,
\]
\[
\varepsilon_{jk}^2=(1-r_{jk}^2)/D,\quad
D=1-r_{12}^2-r_{13}^2-r_{23}^2+2r_{12}r_{13}r_{23}\,.
\]

\begin{figure}
\epsfxsize=9.cm
\centering
\epsfbox{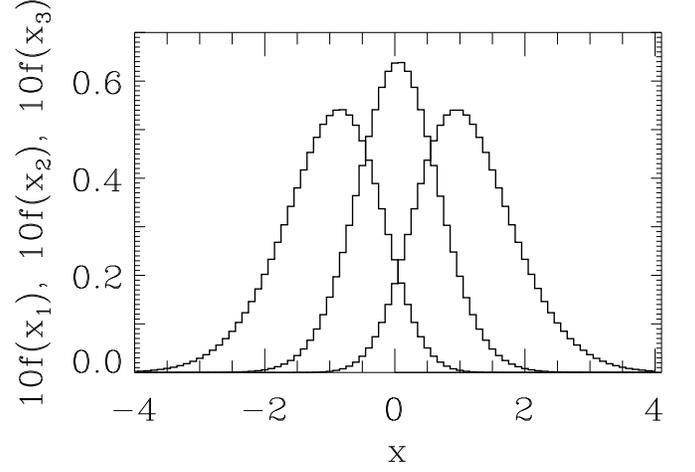}
\vspace{-0.4cm}
\caption{The probability distribution functions for three
displacements $\Delta S_1\geq \Delta S_2\geq \Delta S_3$
are plotted vs. $x=\Delta S_i/\sigma_s$,
$\langle x_i\rangle=1.07, 0, -1.07$,
$\sigma_i=0.76,\,0.64,\,0.76$.
}
\label{pdf}
\end{figure}

As it follows from (\ref{r123}) for the spectra
discussed in Sec. 3\,\&\,4 we get
\[
r_{ij}\simeq 2/3,\quad D \simeq 7/27,\quad
\omega_{jk}^2=15/7,\quad \kappa_{ij}= 2/5\,.
\]
These PDFs of the displacements are plotted in
Fig. \ref{pdf} and some numerical estimates are given in
Table \ref{tbl-frac}.

To convert the expression (\ref{gauss1}) to the orthogonal
form we use transformation
\be
Q=\frac{1}{2}(\eta_1^2+\eta_2^2+\eta_3^2),\quad
\ee
\[
\xi_3=\omega_{33}\eta_3,
\quad \xi_2=\omega_{22}\eta_2-\omega_{23}\eta_3,\quad
\xi_1=\eta_1-\kappa_{12}\xi_2-\kappa_{13}\xi_3\,,
\]
\[
\omega_{33}=\sqrt{\frac{1-\kappa_{12}^2}{DD}}\simeq 1.045,\quad
\omega_{22}=\frac{1}{\sqrt{1-\kappa_{12}^2}}\simeq 1.033\,,
\]
\[
\omega_{23}=\frac{\kappa_{23}-\kappa_{12}\kappa_{13}}{\sqrt{(1-
\kappa_{12}^2)DD}}\simeq 0.21\,,
\]
where
\[
DD=1-\kappa_{12}^2-\kappa_{13}^2-\kappa_{23}^2+
2\kappa_{12}\kappa_{13}\kappa_{23}\simeq 27/32\,.
\]

\end{document}